\documentclass[aps,pre,twocolumn,groupedaddress,showpacs]{revtex4}
\usepackage{graphicx}
\usepackage{bm,bbm,amssymb,dcolumn}
\usepackage{subfigure}

\newcommand{\vect}[1]{\boldsymbol{#1}}
\newcommand{\ten}[1]{\boldsymbol{#1}}
\newcommand{\dif}[1]{\textrm{d} #1 \ }

\newcommand{\difff}[1]{\textrm{d}^3 \vect{#1} \ }
\newcommand{\eq}[1]{\begin{eqnarray}
{#1}
\end{eqnarray}}

\begin{document}

\title{Affine model of stress stiffening in semiflexible filament networks}

\author{J.R.~Blundell and E.M.~Terentjev}

\affiliation{Cavendish Laboratory, University of Cambridge,
Madingley Road, Cambridge, CB3 0HE, U.K. }

\date{\today}

\begin{abstract}

We present a revised theoretical study of the affine assumption applied to semiflexible networks. Drawing on simple models of semiflexible worm-like chains we derive an expression for the probability distribution of crosslink separations valid at all separations. This accounts for both entropic and mechanical filament stretching. From this we obtain the free energy density of such networks explicitly as a function of applied strain. We are therefore able to calculate the elastic moduli of such networks for any imposed strain or stress. We find that accounting for the distribution of cross-link separations destroys the simple scaling of modulus with stress that is well known in single chains, and that such scaling is sensitive to the mechanical stretch modulus of individual filaments. We compare this model to three experimental data sets, for networks of different types of filaments, and find that a properly treated affine model can successfully account for the data. We find that for networks of stiffer filaments, such as F-actin, to fit data we require a much smaller effective persistence length than usually assumed to be characteristic of this filament type. We propose that such an effectively reduced rigidity of filaments might be a consequence of network formation.

\end{abstract}


\maketitle

%%%%%%%%%%%%%%%%%%%%%%%%%%%%%%%%%%%%%%%%%%%%%%%%%%%%%%%%%%%%%%%%%%%%%%%%%%%%%%%%%
\section{Introduction}

Networks of crosslinked or branched semiflexible filaments are found in many biologically relevant systems. These include the networks of filaments such as F-actin, intermediate filaments and microtubules that make up the cytoskeleton of cells, as well as networks of collagen and fibrin found in the extracellular matrix. Looking broader, similar semiflexible/rigid networks are formed by segments of DNA \cite{donald}, by aggregated amyloid fibrils \cite{amyloid}, self-assembled peptide nanotubes \cite{amyloid2} and by carbon nanotubes, on their own or dispersed in polymer matrix \cite{baloo}. Semiflexible networks of cytoskeleton have interesting and unusual elastic properties that are thought to be crucial to the way in which cells move, function and respond to their surroundings \cite{bray, janmey1, guck}. In particular, such networks tend to be stiffer than conventional polymer networks such as rubber, and show a dramatic stiffening over modest strains that is absent in conventional elastomers \cite{mac, weitz, fletcher, janmey}.

This novel behavior is a consequence of the semiflexible nature of filaments that make up the network. Such filaments are much stiffer than conventional polymer chains with persistence lengths $l_p$ ranging from hundreds of nanometers to tens of microns \cite{howard, aebi}. The persistence length is of a similar magnitude to other lengths in the system namely the total filament length $L$ and the characteristic length between crosslinks or branch points $l_c$. As a consequence the filaments cannot be modeled as flexible chains ($l_c \gg l_p$), nor as rigid rods ($l_p \gg l_c$), instead they are termed semiflexible because $l_c \sim l_p$. It is this interplay of three different length scales that distinguishes the elastic behavior of biopolymer networks from rubbers where there is only one relevant length scale in the problem: the span between crosslinking points.

\begin{figure}
\centering
\resizebox{0.35\textwidth}{!}{\includegraphics{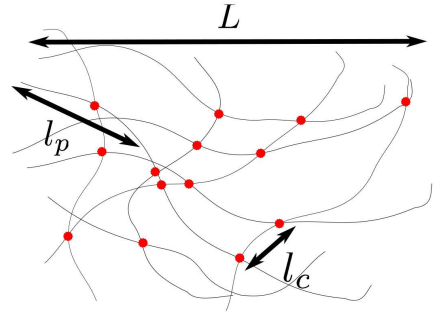}}
 \caption{A schematic illustration of the three relevant length scales in semiflexible networks. Because all three lengths are of a similar scale, some very interesting elastic behavior results.} \label{lengthscales}
\end{figure}

There have been various theoretical ideas put forward to explain the elasticity of these semiflexible networks \cite{mac, weitz, mac0, onck, frey1}. In general these can be split into two classes. Models where the elasticity is explained in terms of the entropic stretching of individual filaments between crosslinks \cite{mac, weitz, mac0} and models where the elasticity is explained in terms of a transition from an enthalpic bending-dominated regime to an entropic stretching dominated regime \cite{frey1, onck}. One of the main distinctions between the two approaches is that for models considering only entropic stretching of filaments it is assumed that any deformation applied to the system is homogeneous down to the length scale of crosslinks, that is, the deformation field is affine \cite{mac}. For models relying on dynamic bending of filaments, the deformation must necessarily be non-affine \cite{onck, frey}.

Recent experiments have claimed that the strain stiffening of such networks is satisfactorily explained in terms of the affine entropic stretching model \cite{mac, weitz}. In particular Storm et al. \cite{mac} concluded that an affine model that accounts for the possible mechanical stretching of filaments is able to explain the elastic response of isotropic fibrogen gels. This is in a curious contrast with Gardel et al. \cite{weitz}, who reported that the elasticity of crosslinked F-actin filaments at various concentrations is explained in terms of entropic effects only, accounting for no mechanical stretching. In this paper we aim to investigate these ideas more closely by comparing an affine model prediction for the shear modulus of semiflexible networks to experimental data of both Storm et al. \cite{mac} (performed on the more flexible fibrogen networks) and also to the data of Gardel et al. \cite{weitz} (performed on the more rigid F-actin networks). We also extend our model to the the case of uniaxial stretching, and compare it with the in-vivo results of Fernandez et al. \cite{fernandez} who measured the stress-stiffening of entire cells under uniaxial strains.

We develop a simple affine model that accounts for both entropic elasticity and the direct mechanical stretching of filaments in a similar way to Storm et al. \cite{mac}. This is achieved by  calculating the probability distribution for the separations of crosslinks based on expressions obtained for semiflexible worm-like chain models \cite{ha, blundell}, which we extend to be valid at all extensions. Our model is therefore able to calculate the free energy density of the network explicitly and therefore we obtain any desired shear (or Young) modulus at a given imposed strain. We find that such a model does indeed explain the stiffening of more flexible biopolymer gels (such as fibrogen) rather well. This is in agreement with the model of Storm et al. \cite{mac}. We also find that the same affine model is able to explain the stress stiffening of crosslinked networks of the stiffer filaments F-actin \cite{weitz}. However, we find that if one properly accounts for the initial distribution of crosslink separations and orientations in the network, there is no longer a universal scaling of modulus with stress for the network as claimed in \cite{weitz}. Moreover we find that the form of the modulus-stress relation becomes sensitive to the mechanical stretch modulus of the individual filaments. We also note a curious feature, that to achieve an agreement with the F-actin data we have to assume that the effective persistence length of the filaments in the network is much smaller than one usually finds in the literature. In discussing this problem, we find that similar effect is known in various other systems, such as amyloid fibril networks, where many authors report a high persistence length on the basis of image analysis of curvature of filaments deposited on a substrate surface \cite{knowles}, while the few studies of bulk networks of these fibrils \cite{amyloid} give very much shorter persistence lengths. We propose that such a reduced persistence length might be a consequence of the filaments being quenched into a network.

%%%%%%%%%%%%%%%%%%%%%%%%%%%%%%%%%%%%%%%%%%%%%%%%%%%%%%%%%%%%
\section{Quenched networks}
\label{sec:prelim}

\begin{figure}
\centering
\resizebox{0.35\textwidth}{!}{\includegraphics{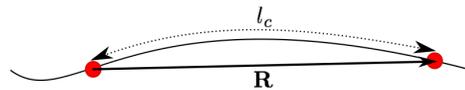}}
 \caption{Schematic illustration of a strand in a semiflexible crosslinked  network. The contour length of filament between the two crosslinks is denoted by $l_c$ while the vector connecting the two crosslinks is denoted by $\vect{R}$.} \label{chain_params}
\end{figure}

The quantities of interest in this paper are the equilibrium elastic moduli of a crosslinked (or branched) semiflexible polymer network. Any such equilibrium thermodynamic property is always expressible as a derivative of the free energy of the system, and so the task is really to be able to calculate the free energy of a crosslinked semiflexible network as a function of arbitrary imposed strain. The remainder of this section is devoted to this task.
Let us initially consider a single section of a filament that connects two crosslink nodes. We will refer to such filament sections as ``strands''. Each strand in the network is identified by the label $\alpha$. The arc length of filament that makes up a strand is denoted by $l_c$ and is assumed to be the same for all $\alpha$ (this is obviously a crude approximation, which however is expected to hold well in a large system with a relatively high crosslinking density). The displacement vector that connects the ends of a strand is denoted by $\vect{R}_{\alpha}$. For the remainder of this paper we will refer to the separation vector as the dimensionless vector $\vect{r} = \vect{R} / l_c$. These variables are shown in Fig. \ref{chain_params}

The vectors $\vect{r}_{\alpha}$ will not all be the same, but will follow some distribution which is a characteristic of the filaments making up the strands. In general, we will label this normalized probability distribution by $P(\vect{r}_{\alpha})$ -- we delay finding an adequate physical expression for $P(\vect{r}_{\alpha})$ until the next section. Given $P(\vect{r}_{\alpha})$, the free energy of the network as a function of strain can be calculated in the following way:

Firstly, we note that the crosslinks are quenched. That is, the time needed for a strand to explore different conformations is much smaller than the time needed for the crosslinks to break and re-form. This separation of timescales means that one can perform equilibrium statistical mechanics over conformations while assuming that the topology of the network remains quenched in place -- a classical procedure of quenched averaging employed in theories of rubber elasticity \cite{flory2, flory, treloar, edwards}.

Secondly, we assume that the strands of the network do not interact apart from at crosslinks (i.e. they are not entangled on their own). This approximation is also well known from rubber elasticity and we expect it to hold for open-mesh semiflexible networks provided $l_c/l_p \lesssim 1$. This is due to the fact that at such length scales the strands behave very rigidly and therefore have less opportunity to explore the surrounding space, and hence we expect excluded volume effects to be negligible. Each strand $\alpha$ will have a free energy of:

\eq{F(\vect{r}_{\alpha}) = k_B T \ln \frac{1}{P(\vect{r}_{\alpha})}  }
If one now deforms the whole network by some strain tensor $\ten{\Lambda}$, the separation vectors $\vect{r}_{\alpha}$ will individually transform to a new separation $\vect{r}'_{\alpha}$ which will depend in some way on the imposed deformation:

\eq{\vect{r}_{\alpha}' = \ten{\lambda}_{\alpha} \vect{r}_{\alpha}}
where $\ten{\lambda}_{\alpha}(\ten{\Lambda})$ is the strain tensor describing the transformation of the $\alpha$th strand. In general the strain tensor $\ten{\lambda}_{\alpha}$ will be a function both of the applied strain $\ten{\Lambda}$ and of which strand we are considering, $\alpha$ -- that is, the deformation does not in general have to be affine.

As long as the topology of the network is conserved under the deformation, the free energy of the network can be expressed as the free energy of each strand in the new strained state $F(\ten{\lambda}_{\alpha} \vect{r}_{\alpha})$ summed over the same strand labels $\alpha$:

\eq{f(\ten{\Lambda}) = \frac{1}{V} \sum_{\alpha} F(\ten{\lambda}_{\alpha} \vect{r}_{\alpha})}

We now make the assumption that the deformation field is affine. That is to say, that the strain tensors acting on each individual strand is the same as the bulk $\ten{\lambda}_{\alpha} = \ten{\Lambda}$. We do not attempt to justify this assumption, but rather will examine the validity of it when comparing the results of this model to experimental results. With this affine assumption the free energy density of a quenched network undergoing affine deformation can be written as the integral:

\eq{f(\ten{\Lambda}) = n k_B T \int \difff{r} P(\vect{r}) \ln \frac{1}{P(\ten{\Lambda}\vect{r})}    \label{free}   }
where $n$ is the number density of strands.
Provided one can obtain an expression for the probability of separation vector $\vect{r}$ occurring, one can calculate the free energy density of the network using the above integral. The next section deals with the details of implementing this calculation: in particular in finding an appropriate expression for $P(\vect{r})$.

%%%%%%%%%%%%%%%%%%%%%%%%%%%%%%%%%%%%%%%%%%%%%%%%%%%%%%%%%%%%%%%%%%%

\section{The model}
\label{sec:model}

This section will be concerned with developing an expression for the probability distribution of semiflexible separation vectors $P(\vect{r})$ that can be used in Eq.\ref{free} to calculate the free energy density of the network. The main difficulty is that although there are a number of expressions that have been proposed for the form of $P(\vect{r})$ \cite{blundell, ha, kleinert, wilhelm},
all are calculated for inextensible chains and therefore have an essential zero at $|\vect{r}| = 1$.

\begin{figure}
\centering
\subfigure[]{ \includegraphics[width=6cm]{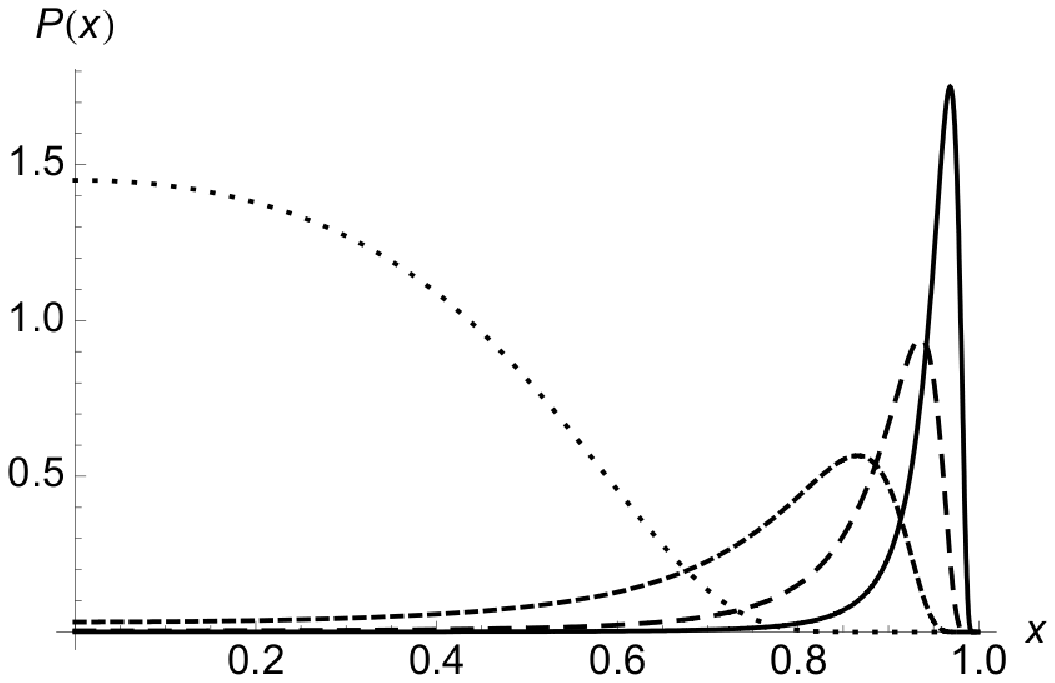}
}
\hspace{1cm}
\subfigure[]{ \includegraphics[width=6cm]{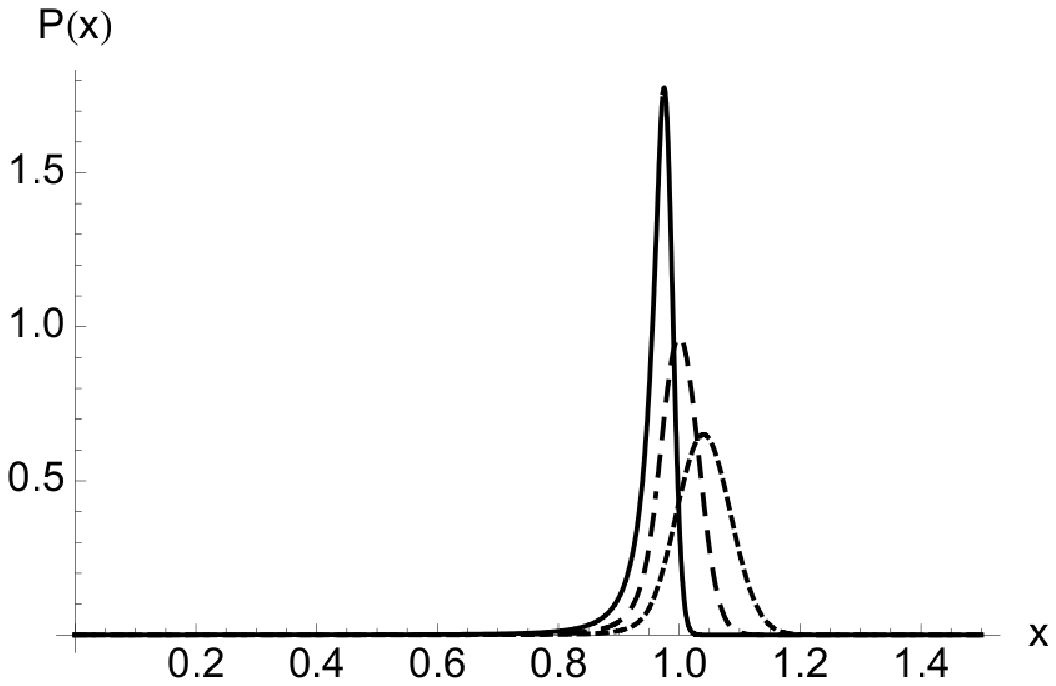}}
\caption{(a)  $P(r)$ plotted for values of $a$ of 0.25 (solid line), 0.5 (large dash), 1.0 (small dash), 5.0 (dots). Note that the last distribution (for more flexible chains) already resembles the Gaussian, with the peak at $r=0$. The value of $\mu_m/ k_B T$ used here is $10^4$. \  (b) $P(r)$ plotted for values of $\mu_m/k_B T$ of 5000 (solid line) 1000 (large dash) and 500 (small dash). The value of $a$ is 0.2} \label{dist1}
\end{figure}

If one invokes the affine assumption however, some chains will inevitably be stretched beyond the limit $|\vect{r}| = 1$. One therefore needs a probability distribution that accounts for the finite probability of the strand being in a mechanically stretched state - where the contour length of the strand has increased beyond $l_c$ by means of stretching of bonds along the filament backbone. The work of Kierfeld et. al. \cite{lipowsky} provides an analysis of single filaments that accounts for the possible mechanical stretching, however they succeed only in providing extension-force relations for such filaments in different force regimes, whereas we require the full probability distribution. We achieve this in the following way:

We again examine a single strand that is at its equilibrium separation. When the separation of the strand ends $\vect{r}$ is at its equilibrium, we expect there to be no backbone stretching. The probability distribution is determined only by entropic effects of the semiflexible filament. In this regime $P(\vect{r})$ is accurately described by distributions obtained in \cite{ha, blundell}, e.g.:

\eq{P({r}) \propto \frac{1}{(1-r^2)^{9/2}}  \textrm{Exp} \left[-\frac{9 a }{8 (1- r^2)}\right]}
where the parameter $a = l_c / l_p$ and is a measure of the flexibility of the filaments and $r = |\vect{r}|$. As one stretches the strand from this equilibrium separation, the entropic returning force of the strand increases and it becomes increasingly difficult to stretch. The effective modulus of this entropic elasticity $\mu_e$ is given by:

\eq{\mu_e(r) = k_B T \partial^2_r \ln \frac{1}{p(r)} }
This is a function of extension $r$ and diverges like $(1-r)^3$ in agreement with \cite{odijk}. At some value of extension, which we will label $r_1$,  the entropic stretching modulus $\mu_e(r)$ will exceed the mechanical stretching modulus of the strand $\mu_m$ (related to the Young modulus of the filament), $\mu_e(r_1, a) = \mu_m$. This defines $r_1$ in terms of $\mu_m$ and $a$. We shall use a simple model for the cross-over from entropic stretching to mechanical stretching, which states that below $r_1$ the response of the chain is purely entropic, while above $r_1$ it is purely mechanical:

\begin{equation}
\mu(r) = \left\{
\begin{array}{ccl}
\mu_e(r) & \ \  0 < r \leq r_1 & \ \ \textrm{Entropic}\\
\mu_m & \ \ r_1 < r \leq \infty & \ \ \textrm{Mechanical}\\
\end{array} \right.
\end{equation}
Note we assume that the Young modulus of the filament remains constant with strain. In reality it would decrease with strain until eventual rupture of the filament, however we do not deal with such terminal effects in this paper. From these expressions for the stretch modulus of an individual strand, we can obtain an expression for the free energy of the strand as a function of separation $r$ by integrating twice:

\begin{equation}
\beta F(r) = \left\{
\begin{array}{cc}
\frac{9}{2} \ln (1-r^2) + \frac{9 a}{8 (1-r^2)}& \ \  0 < r \leq r_1 \\
 \frac{1}{2}\mu_mx^2 +Ax +B  & \ \ r_1 < r \leq \infty \\
\end{array} \right.
\end{equation}
where $A$ and $B$ are constants that are determined by requiring that the free energy and its first derivative with respect to separation are continuous across the boundary at $r_1$.
Given this free energy we can obtain an expression for the probability distribution for the chain that is valid for all separations $r$ using the usual definition:

\eq{P(r) = e^{-\beta [F(r) - F ]},}
where $\beta = 1/k_BT$ and
\eq{F = \int \difff{r} e^{-\beta F(\vect{r})}.}
We remind the reader that the separation is a vector, and therefore integration must be performed over all orientations as well as the length of $\vect{r}$. Plots of $P(r)$ for various values of the parameters $a$ and $\mu_m$ are shown in Fig. \ref{dist1}(a) and (b).

The advantage of our approach is now clear; we have a relatively simple expression for the probability distribution of strand separation vectors that is valid for all separations, from which we can calculate any thermodynamic property related to a single strand. With the probability distribution defined, we can proceed to calculate the free energy density of the network as a function of applied strain via Eq.\ref{free}. This is implemented as follows:

Firstly we must find the initial minimum of the free energy as this is the equilibrium point where the network starts before being strained. As observed in \cite{mac}, this will not in general occur at zero strain. Instead we must allow for the possibility that the network undergoes an initial spherically symmetric bulk volume change (an effect analogous to syneresis in ordinary gels). That is, we allow for the possibility that all strand separations undergo the transformation $|\vect{r}| \rightarrow b |\vect{r}|$,
where $b$ is a constant close to one. Only a spherically symmetric deformation is allowed since by symmetry it is the only spontaneous deformation possible in an isotropic network. We must therefore find the minimum of the integral

\eq{- \int_0^{\infty} 4 \pi r^2 \dif{r} P(r) \ln P(b r)}
with respect to $b$. We solve this numerically. As pointed out in \cite{mac}, $b$ is not an extra parameter in the problem, but is completely determined by $a$ and $\mu_m$.  For networks considered in this paper, all values of $b$ are less than one; the network therefore shrinks initially after crosslinking, which is a well-known effect of reducing the conformational freedom of filaments on establishing quenched crosslinks. Once the minimum of the free energy has been found, we calculate the free energy density as a function of strain using Eq.\ref{free}. We do this for two deformation geometries: shear and uniaxial extension. These are implemented as follows:

\subsection{Simple shear deformation}

The experiments \cite{mac, weitz} impose a simple volume-conserving shear deformation on the networks. Transformation of the separation vector components $(x, y, z)$ under simple shear depends on a single parameter $\gamma$:

\eq{ \left(
\begin{array}{c}
x' \\
y' \\
z'
\end{array} \right) = \left(
\begin{array}{ccc}
1 & 0 & \gamma \\
0 & 1 & 0 \\
0 & 0 & 1
\end{array} \right) \left(
\begin{array}{c}
x \\
y \\
z
\end{array} \right) }

Transforming to spherical polar coordinates we can write the norm of the transformed separation $r'$ in terms of the original separation $r$, the angles parameterizing the orientation of a strand $(\theta, \phi)$ (defined in the usual way with respect to the polar axis $z$) and the applied shear $\gamma$:

\eq{r' = b r \sqrt{1+ 2 \gamma \sin \theta \cos \theta \cos \phi + \gamma^2 \cos^2 \theta}}
The integral to be evaluated is therefore:

\eq{f(\gamma) = - n k_B T \int P(r) \ln P(r') r^2 \sin \theta  {\rm d} r \, {\rm d} \theta \, {\rm d} \phi}

This integral is an expression for the free energy of a semiflexible network that has undergone affine deformation at any shear strain $\gamma$. The integral is performed easily in \textit{Mathematica}. There is sometimes a confusion in the literature about which quantity to call the ``elastic modulus'' when the deformation is not infinitesimal. Two different definitions are possible: the shear modulus $G$ and the differential shear modulus $K$ of the network that has been subject to a simple shear of magnitude $\gamma$ in the $xz$ plane are easily obtained from the free energy density by using the following relations \cite{Landau}:
\eq{G = \frac{\sigma_{xz}}{\gamma} = \frac{2}{\gamma} \frac{\partial f}{\partial \gamma} \qquad K = \frac{\partial \sigma_{xz}}{\partial \gamma} = 2\frac{\partial^2 f}{\partial \gamma^2}}.

\subsection{Uniaxial extension}

There are a number of experiments that impose uniaxial compression or extension on the cytoskeleton, in vitro \cite{fletcher} or in actual living cells \cite{fernandez}. To compare this affine model to such experiments we must calculate the modulus in this geometry. Assuming this deformation is volume conserving, the affine transformation of the separation vector is:

 \eq{ \left(
\begin{array}{c}
x' \\
y' \\
z'
\end{array} \right) = \left(
\begin{array}{ccc}
1/\sqrt{\lambda} & 0 & 0 \\
0 & 1/\sqrt{\lambda} & 0 \\
0 & 0 & \lambda
\end{array} \right) \left(
\begin{array}{c}
x \\
y \\
z
\end{array} \right) }

As before, expressing this in spherical polar coordinates, the norm of the separation vector transforms as:

\eq{r'= b r \sqrt{\frac{\sin^2 \theta}{\lambda} + \lambda^2 \cos^2 \theta  }}

Following the same arguments as in the case of simple shear, the free energy density of a network that has been subject to affine uniaxial volume conserving strain is therefore:

\eq{f(\lambda) = -2 \pi n k_B T \int P(r) \ln P(r') r^2 \sin \theta {\rm d} r \,
{\rm d} \theta , }
where we have picked up a factor of $2 \pi$ from the blank integration over $\phi$. Given this elastic free energy density, the Young modulus $Y$ and differential Young modulus $E$ of the network are given by:

\eq{Y = \frac{\sigma_{zz}}{\lambda} = \frac{1}{\lambda} \frac{\partial f}{\partial \lambda} \qquad E = \frac{\partial \sigma_{zz}}{\partial \lambda} = \frac{\partial^2 f}{\partial \lambda^2}}

\subsection{Scaling of moduli}

For both shear and uniaxial-extension geometries, we can examine how the stiffness of semiflexible networks scales with model parameters. Of particular interest is how the linear modulus of the network (at infinitesimal deformations) scales with the parameter $a = l_c / l_p$, and how the non-linear modulus scales with applied pre-stress. For the affine model considered in this work we find that the linear shear modulus $G_0$ and the linear extension modulus $Y_0$ have a scaling with $a$ of the form:

\eq{G_0 \approx 1.1 n k_BT  \left( \frac{4-a}{a} \right)^2  ; \ \  Y_0 \approx 3.3 n k_BT  \left( \frac{4-a}{a} \right)^2 . }
Naturally, both linear moduli are independent of $\mu_m$ at infinitesimal strains, and they are related by a factor of 3 as indeed required in an incompressible linear elasticity \cite{Landau}. The scaling of $a^{-2}$ is how the ``modulus'' of  single filament scales for large strains. The contribution of the form $(4-a)^2$ comes from the quenched averaging. Once we quench the network with cross-links, we allow a bulk spherically symmetric transformation of all separation vectors, to minimize the quenched free energy. This contribution accounts for the $(4-a)^2$ dependence. Note that for small $a = l_c/l_p$ (very rigid filaments), the scaling is approximately inverse square.

\begin{figure}
\centering
\resizebox{0.35\textwidth}{!}{\includegraphics{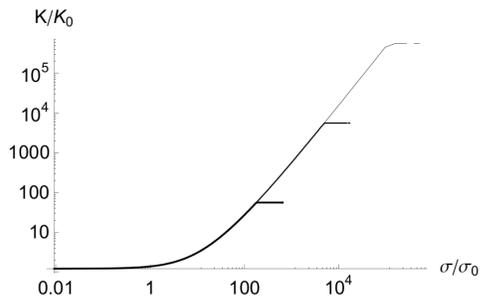}} \caption{Log-log graph of the modulus of a single filament as a function of applied tension, where there is no initial distribution of separation vectors. The curves correspond to $\beta \mu_m = 10^8$, $10^6$ and $10^4$ (top to bottom), for a fixed value of $a=1$. The constant gradient at high tensions corresponds to a power law scaling of $\partial \tau / \partial \lambda \sim \tau^{3/2}$)}
\label{stressscale1}
\end{figure}

\begin{figure}
\centering
\resizebox{0.35\textwidth}{!}{\includegraphics{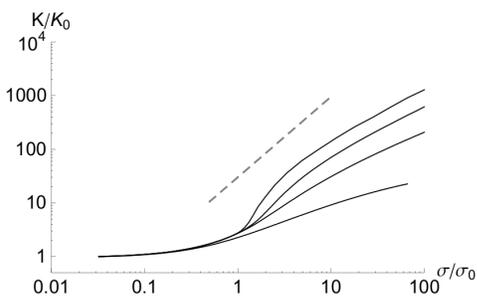}} \caption{Log-log graph of the average modulus as a function of the average tension, where averages are performed over the distribution of separation vectors $P(r)$. Curves are for $\beta \mu_m = 10^{10}$, $10^8$, $10^6$, $10^4$ (top to bottom). Accounting for this distribution destroys the simple power law scaling of modulus with tension, even in a $1-d$ network. The dashed line shows a power law of $3/2$. }
\label{stressscale2}
\end{figure}

It has been claimed that for models considering the simple affine entropic stretching of constituent strands, the scaling of the modulus $K$ with stress $\sigma$ should follow a simple power law $K(\sigma) \sim \sigma^{3/2}$ \cite{weitz}. We find that this reasoning is, however, only valid if one considers a single chain. If instead one properly treats the orientational averaging over the distribution of separation vectors, such universal scaling is lost. We illustrate this effect in Figs. \ref{stressscale1} and \ref{stressscale2} by plotting the stiffness of a single filament as a function of tension, for various values of the stretch modulus $\mu_m$ at fixed $a$. In Fig. \ref{stressscale1}, we plot the ``modulus'' of a single chain $\partial \tau / \partial x$, where $\tau$ is the tension on the filament. As expected, at high $\tau$ the filaments show a power law scaling $ \sim \tau^{3/2}$: it is indeed true that for a worm-like chain model of a single semi-flexible filament, the modulus scales like the tension to the power 1.5 for large extensions, exactly as argued in \cite{weitz} and \cite{fixman}. However, if we have a crosslinked network with a \textit{collection} of filaments with a corresponding distribution of separation vectors $P(r)$, this result changes. In Fig. \ref{stressscale2} we plot the ``modulus'' $\partial \langle \tau \rangle_{P} / \partial \lambda$ as a function of ``stress'' $\langle \tau \rangle_{P}$ for different value of $\mu_m$, where $\langle .. \rangle_{P}$ refers to averages over the probability distribution of separation vectors. We find that no such universal scaling exists -- the functional dependence is no longer a simple power law. In particular, the initial non-linear exponent can be made greater than $3/2$ provided $\mu_m$ is large enough.

%%%%%%%%%%%%%%%%%%%%%%%%%%%%%%%%%%%%%%%%%%%%%%%%%%%%%%%%%%%%%%%%%%%

\section{Results}
\label{sec:results}

We now compare the results of the model presented above with a series of experiments. Firstly we examine the data of Storm et al. \cite{mac}, who measure the simple shear modulus $G$ of branched fibrogen networks as a function of applied strain at various fibrogen concentrations. The data of \cite{mac} is taken over a wide range of strains (up to $\sim 100\%$) and allows the model to be tested well into the regime of non-linear elasticity, up to the point of network rupture. It does not, however, provide a good opportunity to examine the behavior of such networks at low strains ($\leq 5\%$) because the first data point already occurs at $5\%$ strain.

Secondly we compare the predictions of our affine model to the data of Gardel et al. \cite{weitz}, who examine the differential modulus $K$ of crosslinked actin networks, as a function of applied prestress at various concentrations of actin monomer, while keeping the ratio of monomer concentration to crosslinker concentration constant.  This enables us to examine more closely how well an affine model compares to experimental results in stiffer networks at smaller strains.

We also compare our model to the experimental results of Fernandez et al.\cite{fernandez} who measure the uniaxial stiffness of entire living cells as a function of pre-stress. They find that the uniaxial stiffness exhibits a universal scaling with applied pre-stress, which we find is consistent with cells, whose stiffness is determined by an actin cytoskeleton.

\subsection{Comparing to data of Storm et. al.}

\begin{figure}
\centering
\resizebox{0.35\textwidth}{!}{\includegraphics{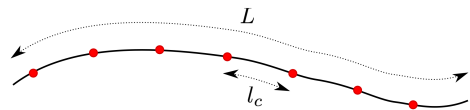}} \caption{Schematic illustration of many strands of length $l_c$ along a filament of length $L$. The approximate number of such strands on a single filament is $m \approx L/l_c$.}
\label{lengths}
\end{figure}

Before we compare our model to experiment we must choose how we are to fit to data. The model we formulated in Eq.\ref{free} has three parameters: $n$ (the number density of strands in the network), $a$ (the ratio of the length of the strand to the filament persistence length $l_c / l_p$) and $\mu_m$ (the mechanical stretching modulus of the filaments). However, the number density of strands $n$ and the value of $a$ are not independent. They are related in the following way: Let us say that we have a filament of total length $L$. Crosslinks or branching points are spaced by an average length $l_c$ along the filament backbone. Assuming $l_c \ll L$, the number of strands $m$ on a single filament is: ${m \approx L/l_c}$. This does not account for dangling ends, but in a densely cross-lined network this becomes irrelevant as $m$ and $m-2$ would differ by a small fraction. These lengths are shown in Fig. \ref{lengths}.

The total number of such filaments in our system must approximately be $N = L_{tot} / L$ where $L_{tot}$ is the total length of all filaments in the system.  This total length of all filaments can be written as the total mass of filament monomer, $M$, divided by the mass per unit length of filament, $\rho$. So the number density of filaments is $c_m /L \rho$, where $c_m$ is the mass density of monomers, $c_m=M/V$. Therefore the total number density of strands in the system is:

\eq{n = \frac{c_m}{L \rho} m = \frac{ c_m}{l_c \rho} = \frac{ c_m}{a l_p \rho} \label{n}}
Since $c_m$, $l_p$ and $\rho$ are independently measured experimentally, the value of $n$ is determined directly by the value of $a$. In examining experimental data, we will fit $a$ and $n$ independently, and then use the above relation as a check on consistency of the fits. We propose that if the fitted parameters $a$ and $n$ are consistent, then the value of $n a l_p \rho / c_m$ should be close to unity.

\begin{figure}
\centering
\subfigure[]{ \includegraphics[width=7cm]{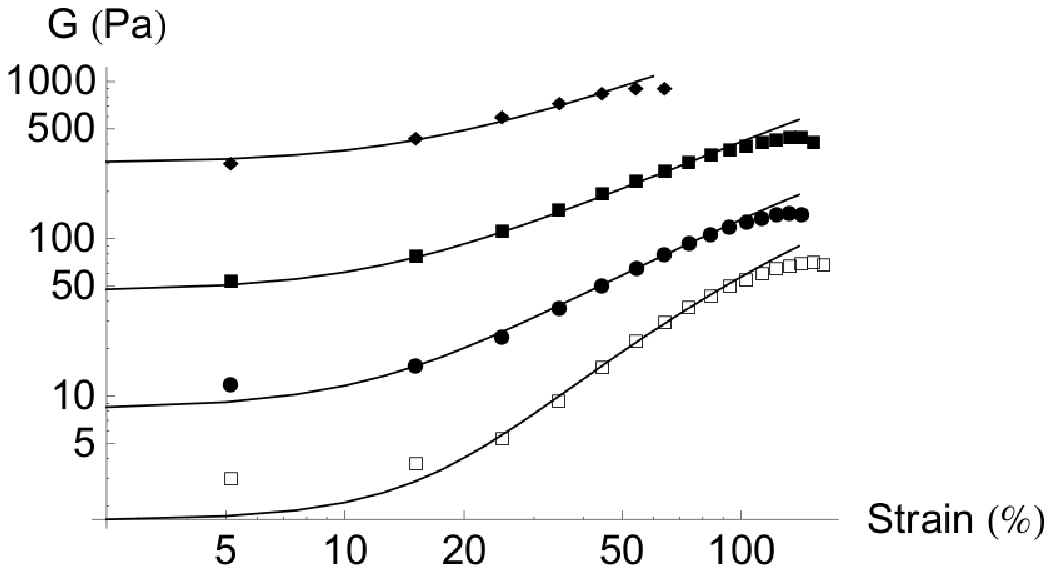}
}
\hspace{1cm}
\subfigure[]{ \includegraphics[width=7cm]{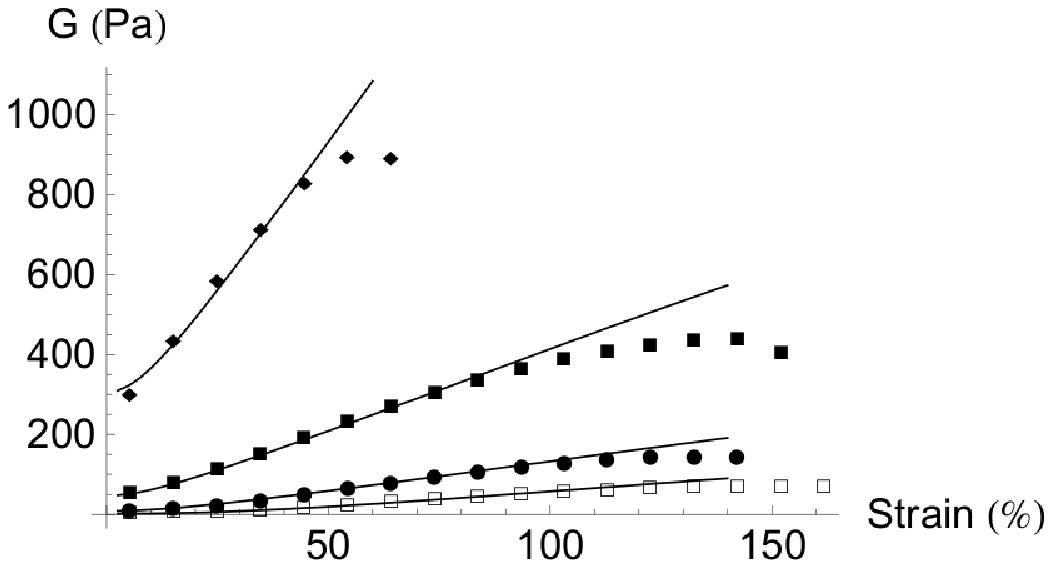}}
\caption{(a) Fit of the model to the data of Storm et al. \cite{mac}. The data sets are obtained from shear measurements on branched fibrogen networks with monomer mass concentrations of $4.5$ ($\blacklozenge$), $2.0$ ($\blacksquare$), $1.0$ ($\bullet$) and $0.5$ ($\square$) mg/ml. The model fits very well to the data at all concentrations for physically reasonable and consistent fit parameters (Table \ref{naturefitvals}), although fails to capture the concave nature of the curves at high strains, thought to be due to filament rupture. \ (b) The same fit as in shown on a Linear-linear scale.} \label{naturefit1}
\end{figure}

The parameter $\mu_m$ is defined above as the energy scale associated with a fractional increase in filament length. We can obtain an estimate of the value of $\mu_m$ for a strand by realizing that the energy of a filament with a Young modulus $E_f$ and cross-sectional area $A$ that has been stretched from a length $l_c$ to $l$ is:

\eq{\mathcal{E} = \frac{1}{2} E_f l_c A (l/l_c - 1)^2 }
Comparing to the original expression for $\mu_m$ we see:

\eq{\mu_m = E_f A l_c = E_f Al_p a}
An estimate for the mechanical stretching modulus can therefore be obtained, provided one knows the Young modulus of the filament and the approximate radius of the filament. In comparing our model to the data of Storm et al. we fix the value of $\mu_m$ to be $5000k_BT a$. We do this because the Young modulus of the filaments should be the same at all concentrations. This is consistent with the measured diameter of $10$nm for fibrogen filaments and a Young modulus of $\sim 1$MPa, which has been measured experimentally for fibrin \cite{falvo}. In that case, only one truly independent fitting parameter remains (the stiffness ratio $a$) and thus the test is rather stringent. The resulting fits are shown in Fig. \ref{naturefit1}(a) and (b), and the fitted values of $a$ are given in Table. (\ref{naturefitvals}).

\begin{table}[t]
\begin{center}
\begin{tabular}{ccccc}
$c_m (mg/ml)$ & $4.5$ & $2.0$ & $1.0$ & $0.5$  \\
\hline
$\mu_m / k_B T a$ & $5000$ & $5000$ & $5000$ & $5000$  \\
$ a $ & $0.60$ & $0.75$ & $0.95$ & $1.40$  \\
$ b $ & $0.994$ & $0.992$ & $0.987$ & $0.968$  \\
$ k_B T n (J/m^3)$ & $9.7$ & $2.2$ & $0.69$ & $0.32$  \\
$n a l_p \rho / c_m$ & $1.89$ & $1.21$ & $0.96$ & $1.31$
\end{tabular}
\end{center}
\label{naturefitvals}
\caption{Parameter values for the fits to data of Storm et al. \cite{mac}. These fit values have been obtained using the persistence length $l_p$ of $0.5\mu$m (reported in \cite{mac}), and a mass per unit length for fibrogen $\rho$ of $2.4 \times 10^{-14}$kg/m \cite{mac}.}
\end{table}
Note that the check of consistency for the fitted values all fall close to the estimate value of $1$. We believe this shows that the affine assumption is valid for the above kinds of networks.

\subsection{Comparing to the data of Gardel et al.}

To compare the affine model proposed here to the data of Gardel et al. \cite{weitz}, we use a slightly altered method to that described above. We do this for the following reason:
Gardel et al. perform measurements of differential shear modulus $K$ on scruin-crosslinked F-actin filaments. They measure both the number concentration of actin monomer in the system $n_A$ as well as the number concentration of scruin crosslinker in the system $n_s$. They then measure the differential shear modulus as a function of pre-stress for four values of $n_A$ while keeping the ratio $\nu= n_s / n_A = 0.03$ constant. Recalling Eq.\ref{n} we have a relationship between the number density of strands in our system $n$ and the contour length of the strands $l_c$: ${n = {c_m}/{l_c \rho }}$,
where $c_m$ is the mass concentration of monomer and $\rho$ is the mass per unit length for F-actin filaments. If such strands are formed at crosslinks, and each crosslink has an average coordination of $z$ (we shall assume $z=4$ here), then the number density of strands $n$ is related to the number density of cross-links $n_{xl}$ via $n = n_{xl} z / 2$. Therefore knowing $n_{xl}$ from experiment, we have an estimate for $l_c$ via:

\eq{l_c = \frac{2 c_m}{\rho z n_{xl}} = \frac{m_A}{2 \rho \nu} \label{l_c} \approx 0.04 \mu\textrm{m},}
where in the second term we have expressed $l_c$ in terms of the mass of a single actin monomer $m_A$ (taken to be 42kD \cite{sackmann1, boal}) and the ratio of crosslinker concentration to monomer concentration $\nu = n_s / n_A$ which was kept at a constant value of 0.03 in these experiments. In the final term we have evaluated the contour length of a strand for the experimental values of \cite{weitz} and $\rho = 2.6 \times 10^{-14}$kg/m \cite{boal}. Note that this predicts that $l_c$ is only dependent on the ratio of crosslinker to actin monomer $\nu=n_s / n_A$ and so is constant for the different data sets of Gardel et al. This is what we should expect: if strands are only formed at crosslinks as we suppose, then doubling both the number of crosslinks and the number of monomers should not result in a change of $l_c$. This is in contrast with the argument offered in \cite{weitz}, where the authors propose that $l_c$ is proportional to the entanglement length (in turn, correctly estimated in \cite{semenov}) at a fixed $\nu$.

Having an estimate for $l_c$ we now estimate the value of the mechanical stretching modulus $\mu_m$ from the expression ${\mu_m = E_f A l_c}$,
where $E_f$ is the Young modulus of F-actin filaments and $A$ their effective cross-sectional area. We can estimate the value of $\mu_m$ for such filaments from the value for $E_f A = 5 \times 10^{-8} $N measured for F-actin in \cite{kojima}. This yields ${\mu_m = 6 \times 10^5 k_B T}$.
Since the value of $l_p$ is constant for the data sets, the value of $\mu_m$ must also be constant. If we now take a value for the persistence length of actin of $10\mu$m which is most frequently quoted in the literature  \cite{howard, goldstein, janmey2}, we then have estimates for the three parameters of our model ($n$, $\mu_m$, $a$) taken directly from experimentally measured quantities. These are shown in Table \ref{measured}.

\begin{table}[t]
\begin{center}
\begin{tabular}{ccccc}
$c_m (mg/ml)$ & $1.2$ & $0.90$ & $0.50$ & $0.35$  \\
\hline
$\mu_m / k_B T$ & $6 \times10^5$ & $6 \times10^5$ & $6 \times10^5$ & $6 \times10^5$  \\
$ a $ & $0.005$ & $0.005$ & $0.005$ & $0.005$  \\
$ k_B T n (J/m^3)$ & $4.4$ & $3.2$ & $1.8$ & $1.2$  \\
\end{tabular}
\end{center}
\caption{A table of the model parameters we would expect if we take the persistence length of filaments to be $l_p = 10\mu$m. Such a value for $l_p$ would result in very stiff filaments. From Fig. \ref{scifit1} (a) we see that such model parameters produce $K(\sigma)$ curves that drastically fail to agree with experimental data. }
\label{measured}
\end{table}

\begin{figure}
\centering
\subfigure[]{ \includegraphics[width=7cm]{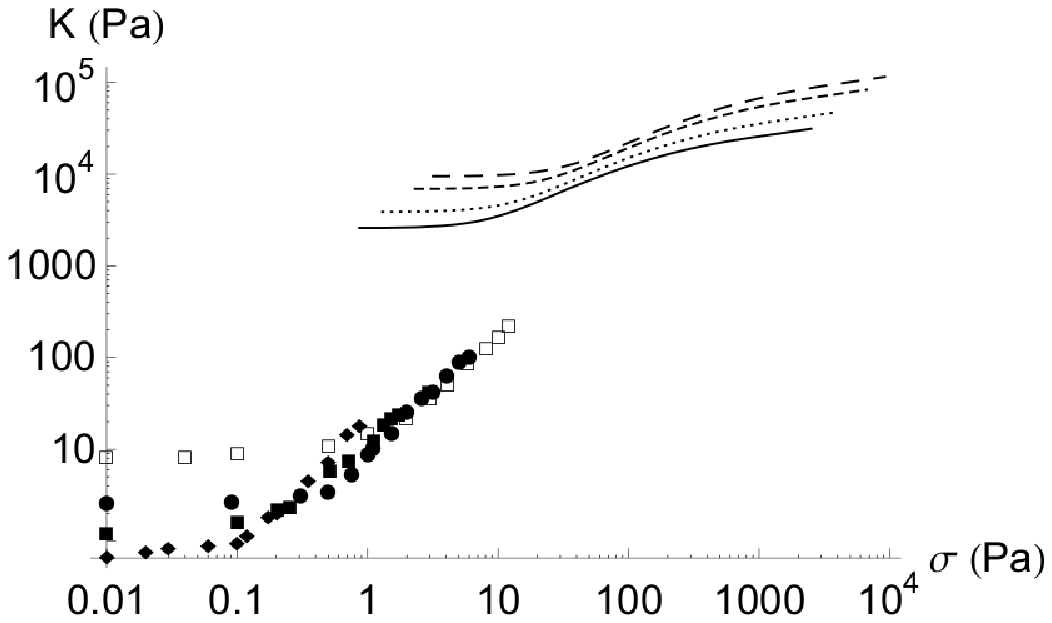}
}
\hspace{1cm}
\subfigure[]{ \includegraphics[width=7cm]{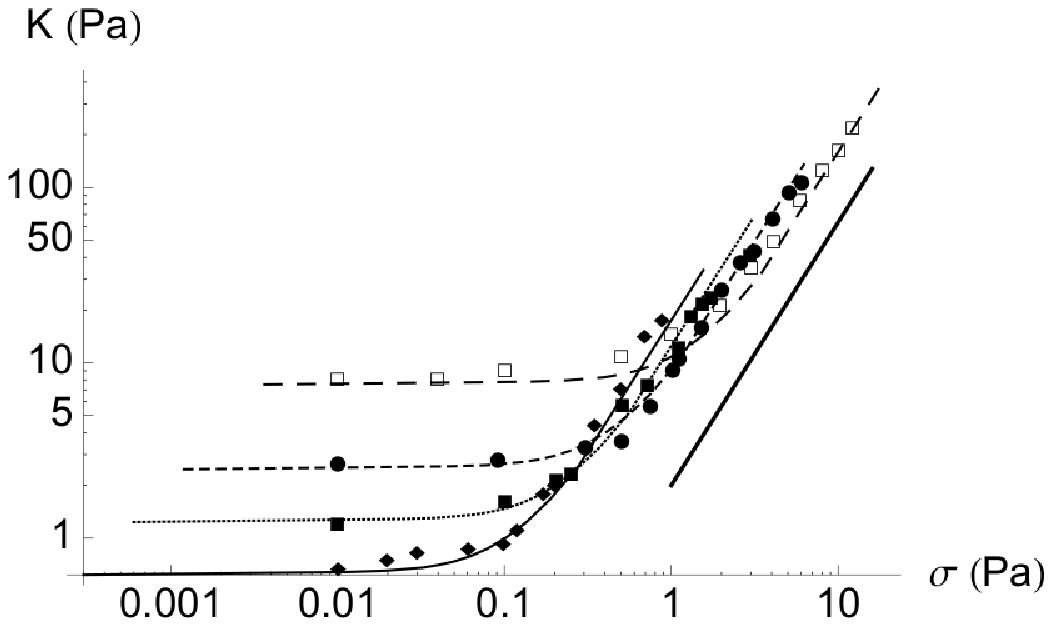}}
\caption{(a)  $K(\sigma)$ predicted by the affine model for the parameters in Table \ref{measured} compared to the experimental data. If one assumes that $l_p = 10\mu$m, then the affine model produces very poor agreement with the data. Apart from the disagreement in values, the model does not show the scaling of $K(\sigma) \sim \sigma^{3/2}$ observed. \ (b) Two-parameter fits of the model to the data of Gardel et al. \cite{weitz} on log-log axes. The data shown corresponds to  differential shear modulus for networks with F-actin monomer concentrations $c_A$ of $29.4$ ($\square$), $21.4$ ($\bullet$), $11.9$ ($\blacksquare$) and $8.33\mu$M ($\blacklozenge$) and constant value of $\nu = 0.03$. The fitted value for $l_p = 0.022\mu$m, assuming the value of $l_c$ remains fixed at $0.04\mu$m. The solid straight line shows a scaling of $\sigma^{3/2}$.} \label{scifit1}
\end{figure}

The predicted curves of differential modulus against applied pre-stress for the parameter values in Table \ref{measured} are shown in Fig. \ref{scifit1} (a) along with the plots of the original data. It is clear such parameter values give very poor fits to data predicting an unstrained modulus $K(0)$ that is out three orders of magnitude.

We would like to point out that in order to get a good fit to experimental data it is not simply a case of rescaling the number density of filament strands in the network $n$. Indeed doing so gives curves that fail to predict the onset of stress stiffening by over an order of magnitude. Remarkably however, we do find that the affine model can provide very good fits to the data of \cite{weitz} as shown in Fig. \ref{scifit1} (b), for the different set of fitting parameters in Table. \ref{scivals}. There are a few interesting points to note. We have fixed the value of both $\mu_m$ and $l_c$ for such fits as we have estimates for both, based on experimentally measurable quantities. We then perform a two-parameter fit by varying $l_p$ and $n$. Good fits can only be achieved with this affine model by using a value for the ratio $a=l_c/l_p$, that is two orders of magnitude larger than would be expected from the simple arguments given above -- that is, a value of $l_p = 0.022\mu$m. As before, we check the consistency of the values for $n$ by calculating $n l_c \rho / c_m$ which should be close to unity -- and indeed we find values that are reasonably consistent. Note we do observe a scaling resembling $K(\sigma) \sim \sigma^{3/2}$, in agreement with the statement in \cite{weitz}. However, we find that such scaling is not universal for such networks but depends strongly on the values of $a= l_c / l_p$ and $\mu_m$. It is in fact because of the particular fitted values of $a$ and $\mu_m$, that we observe a scaling like $K(\sigma) \sim \sigma^{3/2}$.

In short, we find that assuming the persistence length of F-actin filaments in these networks is $\sim 10\mu$m results in the affine model failing severely and predicting that the networks would be far stiffer than is actually measured. More importantly, there is no easy way out of this disagreement because experimental data have several key features that are all linked. However, we find the affine model of filament network can be reconciled with experiment, and produce very good agreement provided we assume that the F-actin filaments behave as if they are far more flexible than is usually quoted -- with a persistence length of $\sim 0.02\mu$m.  Interestingly such behavior has actually been reported experimentally in networks of actin filaments, who report values of the persistence length as low as $0.5\mu$m \cite{sackmann1, sackmann2} which, although still much larger than the value needed by the affine model, shows that the persistence length of filaments embedded in networks can be far less than what is most frequently reported for single actin filaments. It is, of course, also possible that the in-vitro simple shear experimental geometry at large pre-stress underestimates the actual modulus, so that $l_p$ closer to $0.5\mu$m would actually be acceptable.

\begin{table}[t]
\begin{center}
\begin{tabular}{ccccc}
$c_m (mg/ml)$ & $1.2$ & $0.90$ & $0.50$ & $0.35$  \\
\hline
$\mu_m / k_B T$ & $6 \times 10^5$ & $6 \times 10^5$ & $6 \times 10^5$ & $6 \times 10^5$  \\
$ a $ & $2.0$ & $2.0$ & $2.0$ & $2.0$  \\
$ b $ & $0.92$ & $0.92$ & $0.92$ & $0.92$  \\
$ k_B T n $ & $3.75$ & $1.25$ & $0.52$ & $0.29$  \\
$n l_c \lambda / c_m$& $0.89$ & $0.40$ & $0.30$ & $0.24$
\end{tabular}
\end{center}
\caption{Model parameters for the fit to the data of Gardel et al. \cite{weitz} shown in Fig. \ref{scifit1} (b). It is a two-parameter fit for the values of $n$ and $a$. Fitted values are obtained assuming the contour length of  $l_c=0.044\mu$m and a mass per unit length for F-actin $\rho = 2.6 \times 10^{-14}$kg/m. The fitted persistence length is therefore $0.022\mu$m. Provided we take the persistence length to be so small, we achieve good fits to data with reasonably consistent values for $n$. }
\label{scivals}
\end{table}

\subsection{Comparing to the results of Fernandez et. al.}

\begin{figure}
\centering
\resizebox{0.35\textwidth}{!}{\includegraphics{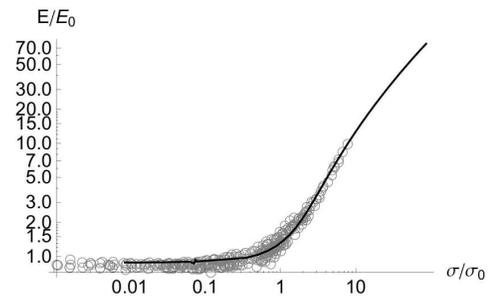}} \caption{Fit of the model to the collapsed data of Fernandez et. al. \cite{fernandez}. A very good fit to data can be achieved by the parameters $a = 0.5$ and $\beta \mu_m = 5 \times 10^6$. These values are consistent with an actin network.}
\label{fernandezfit}
\end{figure}

To highlight the fact that this model might well be of biological significance, we compare its predictions with the data of Fernandez et al. \cite{fernandez} who measure the uniaxial stiffness $E$ of entire cells as a function of applied pre-stress. They have found that the stiffness of a cell follows a simple master equation:

\begin{equation}
\mu(r) = \left\{
\begin{array}{cc}
E = E_0 & \sigma < \sigma_c \\
E \propto \sigma& \sigma > \sigma_c\\
\end{array} \right.
\end{equation}
which is to say, a collapse of the modulus stress curves is achieved by rescaling the modulus axis by $E_0$, and rescaling the stress axis by $\sigma_c$. The model presented here provides a possible explanation for this behavior. The non-linear behavior is a consequence of chains being initially entropically and then subsequently mechanically stretched. The scaling behavior of $E(\sigma)$ is a function of our key parameters $a$ and $\mu_m$, that is, it depends on the transition from entropic stretching to mechanical stretching. We fit this data by for a value of $a = 0.5$ and $\mu_m = 5 \times 10^6 k_BT$, with the result shown in Fig. \ref{fernandezfit}. The value of the fitted parameter $\beta \mu_m = 5 \times 10^6$ is consistent with filaments with Young modulus $E_f \sim 10^9$ Pa, diameters of $d \sim 5nm$ and lengths $l_c \sim 0.5\mu$m. These values are all consistent with an actin network. This suggests that the stiffness of such cells could be determined by the actin cytoskeleton network, with strand lengths of $\sim 0.5 \mu$m and persistence lengths of $\sim 1 \mu$m (note that this is also much smaller than the `accepted' values of $l_p \sim 10 \mu$m).

%%%%%%%%%%%%%%%%%%%%%%%%%%%%%%%%%%%%%%%%%%%%%%%%%%%%%%%%%%%%%%%%%%%

\section{Conclusions}
\label{sec:conc}

In this work we have shown that the equilibrium shear modulus of branched fibrogen networks and crosslinked F-actin networks as a function of applied strain is satisfactorily explained by an affine model that accounts for mechanical as well as entropic stretching. In the case of fibrogen networks we find the affine model fits the data remarkably well with fit parameters which are both consistent with each other and with experimental measurements on individual fibrogen fibers. In the case of F-actin networks we find that the model predicts the observed stress stiffening behavior accurately provided we allow the effective persistence length of the filaments in the network to be more flexible than is expected. That is to say, an affine model agrees with the data well and has reasonably consistent fit parameters provided the effective persistence length of filaments in the network is taken to be $l_p \sim 0.02\mu$m. This is far lower than the observed persistence length of individual F-actin filaments reported in individual measurements \cite{janmey, goldstein}, but a reduced persistence length ($l_p \sim 0.5 \, \mu$m) has been reported experimentally in some F-actin networks \cite{sackmann1, sackmann2}, and also arises from the fitting of the in-vivo data of Fernandez et al. \cite{fernandez}.

This suggests to us that an affine model does rather well at explaining the elasticity of branched or crosslinked semi-flexible filaments because in such networks the filaments behave far more flexibly than expected on the basis of measurements of filament curvature, when they are adsorbed on a substrate (note that the few measurements which measure $l_p$ with quasi-elastic light scattering of filaments in networks often report much shorter $l_p$ \cite{sackmann1, sackmann2}). Because the length of filament strands, $l_c$, is actually of a similar order or even greater than the effective persistence length of the filaments, treating the strands as independent is a reasonable approximation -- as it is for rubber -- where one does indeed expect the affine model to work well. For more rigid networks that are composed of filaments where the effective persistence length is smaller than the contour length of strands, $l_p \gg l_c$, we would expect the affine assumption not to be valid, and models incorporating non-affine deformations may well need to be considered \cite{onck, frey, storm}. However, it appears that in-vitro actin networks extensively studied in the recent years are not in this regime.

We also find that properly accounting for the distribution of cross-link separation vectors has a profound effect on the scaling of the modulus of the network with stress. Averaging over the distribution of initial separation vectors means we do not see a universal scaling of the modulus with stress of the form $K(\sigma) \sim \sigma^{3/2}$ as predicted by \cite{weitz},  but rather a functional dependence that is not a power law, and whose form depends on the values of $a=l_c / l_p$ and $\mu_m$ for filaments in the network. The fact that such a scaling is non-universal, but depends on chain parameters, can be used to explain why different scaling regimes are observed in different networks. Although Gardel et. al. \cite{weitz} observe scaling like $K(\sigma) \sim \sigma^{3/2}$, experiments on dendritic actin \cite{fletcher} have observed a scaling like $E(\sigma) \sim \sigma^{0.3}$, while experiments on entire cells \cite{fernandez} have observed a scaling like $E(\sigma) \sim \sigma$. We have shown that an affine model is able to capture the different scaling behaviors observed by accounting for the mechanical stretching of filaments. This could be of biological significance, as the non-linear behavior of the modulus will in general depend on the parameters of the constituent filaments. This suggests that one can tailor the form of the non-linear behavior of such networks by tuning the filament parameters $a = l_c / l_p$ and $\mu_m = E_f Al_c$. That is, the  form of the non-linear modulus as a function of stress $K(\sigma)$ can be drastically altered by changing $l_c$.

%%%%%%%%%%%%%%%%%%%%%%%%%%%%%%%%%%%%%%%%%%%%%%%%%%%%%%%%%%%%%%%%%%%
\section{Acknowledgements}

It is a pleasure to acknowledge P. Fernanazez and C. Storm for helpful discussions and for providing the data for Fig. \ref{naturefit1} and Fig. \ref{fernandezfit}. This work has been supported by EPSRC funding.

%%%%%%%%%%%%%%%%%%%%%%%%%%%%%%%%%%%%%%%%%%%%%%%%%%%%%%%%%%%%%%%%%%%

%\bibliography{Refs}

\begin{thebibliography}{40}
\providecommand{\url}[1]{\texttt{#1}}
\providecommand{\urlprefix}{ }

\bibitem[Hasnain and Donald(2006)]{donald}
Hasnain, I., and A.~Donald, 2006.
\newblock {Microrheological characterization of anisotropic materials }.
\newblock \emph{Phys. Rev. E} 73:031901.

\bibitem[MacRaild et~al.(2004)MacRaild, Stewart, Mok, Gunzburg, Perugini,
  Lawrence, Tirtaatmadja, Cooper-White, and Howlett]{amyloid}
MacRaild, C., C.~Stewart, Y.-F. Mok, M.~Gunzburg, M.~Perugini, L.~Lawrence,
  V.~Tirtaatmadja, J.~Cooper-White, and G.~Howlett, 2004.
\newblock {Non-fibrillar components of amyloid deposits mediate the
  self-association and tangling of amyloid fibrils }.
\newblock \emph{J. Biol. Chem.} 279:21038.

\bibitem[Mahler et~al.(2006)Mahler, Reches, Rechter, Cohen, and
  Gazit]{amyloid2}
Mahler, A., M.~Reches, M.~Rechter, S.~Cohen, and E.~Gazit, 2006.
\newblock {Rigid, self-assembled hydrogel composed of a modified aromatic
  dipeptide }.
\newblock \emph{Adb. Mater.} 18:1365.

\bibitem[Ahir et~al.(2007)Ahir, Terentjev, Lu, and Panchapakesan]{baloo}
Ahir, S., E.~Terentjev, S.~Lu, and B.~Panchapakesan, 2007.
\newblock {Thermal fluctuations, stress relaxation, and actuation in carbon
  nanotube networks }.
\newblock \emph{Phys. Rev. B} 76:165437.

\bibitem[Alberts et~al.(1994)Alberts, Bray, Lewis, Raff, Roberts, and
  Watson]{bray}
Alberts, B., D.~Bray, J.~Lewis, M.~Raff, K.~Roberts, and J.~Watson, 1994.
\newblock Molecular Biology of the Cell.
\newblock Garland, New York, 3rd edition.

\bibitem[Janmey et~al.(1990)Janmey, Hvidt, Lamb, and Stossel]{janmey1}
Janmey, P., S.~Hvidt, J.~Lamb, and T.~Stossel, 1990.
\newblock Resemblance of actin-binding protein/actin gels to covalently
  crosslinked networks.
\newblock \emph{Nature} 345:89.

\bibitem[Guck et~al.({2005})Guck, Schinkinger, Lincoln, Wottawah, Ebert,
  Romeyke, Lenz, Erickson, Ananthakrishnan, Mitchell, Kas, Ulvick, and
  Bilby]{guck}
Guck, J., S.~Schinkinger, B.~Lincoln, F.~Wottawah, S.~Ebert, M.~Romeyke,
  D.~Lenz, H.~Erickson, R.~Ananthakrishnan, D.~Mitchell, J.~Kas, S.~Ulvick, and
  C.~Bilby, {2005}.
\newblock {Optical deformability as an inherent cell marker for testing
  malignant transformation and metastatic competence}.
\newblock \emph{{Biophys. J.}} {88}:{3689--3698}.

\bibitem[Storm et~al.(2005)Storm, Pastore, MacKintosh, Lubensky, and
  Janmey]{mac}
Storm, C., J.~Pastore, F.~MacKintosh, T.~Lubensky, and P.~Janmey, 2005.
\newblock Non-linear Elasticity in biological gels.
\newblock \emph{Nature} 435:191.

\bibitem[Gardel et~al.(2004)Gardel, Shin, Mackintosh, Mahadevan, Matsudaira,
  and Weitz]{weitz}
Gardel, M., J.~Shin, F.~Mackintosh, L.~Mahadevan, P.~Matsudaira, and D.~Weitz,
  2004.
\newblock Elastic behavior of cross-linked and bundled actin networks.
\newblock \emph{Science} 304:5675.

\bibitem[Chaudhuri et~al.(2005)Chaudhuri, Parekh, and Fletcher]{fletcher}
Chaudhuri, O., S.~H. Parekh, and D.~A. Fletcher, 2005.
\newblock Reversible stress softening of actin networks.
\newblock \emph{Nature} 445:295.

\bibitem[Wen et~al.(2007)Wen, Basu, Winer, Yodh, and Janmey]{janmey}
Wen, Q., A.~Basu, J.~Winer, A.~Yodh, and P.~Janmey, 2007.
\newblock Local and global deformations in a strain-stiffening Þbrin gel.
\newblock \emph{New Journal of Physics} 9:428.

\bibitem[Gittes et~al.({1993})Gittes, Mickey, Nettleton, and Howard]{howard}
Gittes, F., B.~Mickey, J.~Nettleton, and J.~Howard, {1993}.
\newblock {Flexural rigidity of microtubules and actin-filaments measured from
  thermal fluctuations in shape}.
\newblock \emph{{J. Cell. Biol.}} {120}:{923--934}.

\bibitem[Kreplak et~al.({2005})Kreplak, Bar, Leterrier, Herrmann, and
  Aebi]{aebi}
Kreplak, L., H.~Bar, J.~Leterrier, H.~Herrmann, and U.~Aebi, {2005}.
\newblock {Exploring the mechanical behavior of single intermediate filaments}.
\newblock \emph{{J. Mol. Biol.}} {354}:{569--577}.

\bibitem[MacKintosh et~al.(1995)MacKintosh, Kas, and Janmey]{mac0}
MacKintosh, F., J.~Kas, and P.~Janmey, 1995.
\newblock Elasticity of Semiflexible Biopolymer Networks.
\newblock \emph{Phys. Rev. Lett.} 75:4425.

\bibitem[Onck et~al.(2005)Onck, Koeman, van Dillen, and van~der Giessen]{onck}
Onck, P.~R., T.~Koeman, T.~van Dillen, and E.~van~der Giessen, 2005.
\newblock Alternative Explanation of Stiffening in Cross-Linked Semißexible
  Networks.
\newblock \emph{Phys. Rev. Lett.} 95:178102.

\bibitem[Heussinger et~al.({2007})Heussinger, Schaefer, and Frey]{frey1}
Heussinger, C., B.~Schaefer, and E.~Frey, {2007}.
\newblock {Nonaffine rubber elasticity for stiff polymer networks}.
\newblock \emph{{Phys. Rev. E}} {76}.

\bibitem[Heussinger et~al.(2007)Heussinger, Schaefer, and Frey]{frey}
Heussinger, C., B.~Schaefer, and E.~Frey, 2007.
\newblock NonafÞne rubber elasticity for stiff polymer networks.
\newblock \emph{Phys. Rev. E} 76:031906.

\bibitem[Fernandez et~al.({2006})Fernandez, Pullarkat, and Ott]{fernandez}
Fernandez, P., P.~Pullarkat, and A.~Ott, {2006}.
\newblock {A master relation defines the nonlinear viscoelasticity of single
  fibroblasts}.
\newblock \emph{{Biophys. J.}} {90}:{3796--3805}.

\bibitem[Ha and Thirumalai(1997)]{ha}
Ha, B.-Y., and D.~Thirumalai, 1997.
\newblock Semiflexible chains under tension.
\newblock \emph{J. Chem. Phys} 106:4243.

\bibitem[Blundell and Terentjev(2007)]{blundell}
Blundell, J., and E.~Terentjev, 2007.
\newblock Forces and extensions in eemiflexible and rigid polymer chains and
  filaments.
\newblock \emph{J. Phys. A: Math. Theor.} 40:10951.

\bibitem[Knowles et~al.(2006)Knowles, Smith, Craig, Dobson, and
  Welland]{knowles}
Knowles, T., J.~Smith, A.~Craig, C.~Dobson, and M.~Welland, 2006.
\newblock {Spatial persistence of angular correlations in amyloid fibrils }.
\newblock \emph{Phys. Rev. Lett.} 98:238301.

\bibitem[Flory({1943})]{flory2}
Flory, P., {1943}.
\newblock {Statistical mechanics of cross-linked polymer networks I Rubberlike
  elasticity}.
\newblock \emph{{J. Chem. Phys.}} {11}:{512}.

\bibitem[Flory(1969)]{flory}
Flory, P., 1969.
\newblock Statistical Mechanics of Chain Molecules.
\newblock Interscience, New York.

\bibitem[Treloar(2005)]{treloar}
Treloar, L. R.~G., 2005.
\newblock The Physics of Rubber Elasticity.
\newblock OUP, Oxford, 3rd edition.

\bibitem[Deam and Edwards(1976)]{edwards}
Deam, R., and S.~Edwards, 1976.
\newblock Theory of Rubber Elasticity.
\newblock \emph{Phil. Trans. R. Soc. Lon. A} 280:317.

\bibitem[Hamprecht and Kleinert(2005)]{kleinert}
Hamprecht, B., and H.~Kleinert, 2005.
\newblock End-to-end distribution function of stiff polymers for all
  persistence lengths.
\newblock \emph{Phys. Rev. E} 71:031803.

\bibitem[Wilhelm and Frey(1996)]{wilhelm}
Wilhelm, J., and E.~Frey, 1996.
\newblock Radial distribution function for semiflexible polymers.
\newblock \emph{Phys. Rev. Lett.} 77:2581.

\bibitem[Kierfeld et~al.(2004)Kierfeld, Niamploy, Sa-yakanit, and
  Lipowsky]{lipowsky}
Kierfeld, J., O.~Niamploy, V.~Sa-yakanit, and R.~Lipowsky, 2004.
\newblock Stretching of semiflexible polymers with elastic bonds.
\newblock \emph{Eur. Phys. J. E} 14:17.

\bibitem[Odijk(1995)]{odijk}
Odijk, T., 1995.
\newblock Stiff chains and filaments under tension.
\newblock \emph{Macromolecules} 28:7016.

\bibitem[Landau and Lifshitz(1986)]{Landau}
Landau, L., and E.~Lifshitz, 1986.
\newblock Theory of Elasticity.
\newblock Butterworth Heinemann, 3rd edition.

\bibitem[Fixman and Kovac(1973)]{fixman}
Fixman, M., and J.~Kovac, 1973.
\newblock {Polymer conformational statistics .3. Modified Gaussian model of
  stiff chains}.
\newblock \emph{J. Chem. Phys.} 58:1564.

\bibitem[Guthold et~al.({2007})Guthold, Liu, Sparks, Jawerth, Peng, Falvo,
  Superfine, Hantgan, and Lord]{falvo}
Guthold, M., W.~Liu, E.~A. Sparks, L.~M. Jawerth, L.~Peng, M.~Falvo,
  R.~Superfine, R.~R. Hantgan, and S.~T. Lord, {2007}.
\newblock {A comparison of the mechanical and structural properties of fibrin
  fibers with other protein fibers}.
\newblock \emph{{Cell Biochem. Biophys.}} {49}:{165--181}.

\bibitem[Schmidt et~al.({1989})Schmidt, Barmann, Isenberg, and
  Sackmann]{sackmann1}
Schmidt, C.~F., M.~Barmann, G.~Isenberg, and E.~Sackmann, {1989}.
\newblock {Chain Dynamics, Mesh Size, and Diffusive Transport in Networks of
  Polymerized Actin. A Quasielastic Light Scattering and Microfluorescence
  Study}.
\newblock \emph{{Macromolecules }} {22}:{3638}.

\bibitem[Boal(2002)]{boal}
Boal, D., 2002.
\newblock Mechanics of the Cell.
\newblock CUP, Cambridge, 1st edition.

\bibitem[Semenov(1986)]{semenov}
Semenov, A.~N., 1986.
\newblock Dynamics of Concentrated Solutions of Rigid-chain Polymers Part 1
  -Brownian Motion of Persistent Macromolecules in Isotropic Solution.
\newblock \emph{J. Chem Soc. Faraday} 82:317.

\bibitem[Kojima et~al.({1994})Kojima, Ishijima, and Yanagida]{kojima}
Kojima, H., A.~Ishijima, and T.~Yanagida, {1994}.
\newblock {Direct measurement of stiffness of single actin-filaments with and
  without tropomyosin by in-vitro nanomanipulation}.
\newblock \emph{{PNAS}} {91}:{12962}.

\bibitem[Riveline et~al.({1997})Riveline, Wiggins, Goldstein, and
  Ott]{goldstein}
Riveline, D., C.~Wiggins, R.~Goldstein, and A.~Ott, {1997}.
\newblock {Elastohydrodynamic study of actin filaments using fluorescence
  microscopy}.
\newblock \emph{{Phys. Rev. E}} {56}:{1330--1333}.

\bibitem[Janmey et~al.({1986})Janmey, Peetermans, Zaner, Stossel, and
  Tanaka]{janmey2}
Janmey, P., J.~Peetermans, K.~Zaner, T.~Stossel, and T.~Tanaka, {1986}.
\newblock {Structure and mobility of actin-filaments as measured by
  quasi-elastic light scattering, viscometry, and electron microscopy}.
\newblock \emph{{J. Bio. Chem.}} {261}:{8357--8362}.

\bibitem[Piekenbrock and Sackmann({1992})]{sackmann2}
Piekenbrock, T., and E.~Sackmann, {1992}.
\newblock {Quasi-elastic light scattering study of thermal excitations of
  F-actin solutions and of growth kinetics of actin-filaments}.
\newblock \emph{{Biopolymers}} {32}:{1471--1489}.

\bibitem[Huisman et~al.({2008})Huisman, Storm, and Barkema]{storm}
Huisman, E., C.~Storm, and G.~Barkema, {2008}.
\newblock {Monte Carlo study of multiply crosslinked semißexible polymer
  networks}.
\newblock \emph{{Preprint by correspondence arXiv: 0807.0720v1 [cond-mat.soft]
  }} .

\end{thebibliography}
%\input{bibliography2.bbl}

%%%%%%%%%%%%%%%%%%%%%%%%%%%%%%%%%%%%%%%%%%%%%%%%%%%%%%%%%%%%%%%%%%%
\end{document}